\documentclass{acm_proc_article-sp}
\usepackage{graphicx}

\begin{document}

\conferenceinfo{}{Bloomberg Data for Good Exchange 2016, NY, USA}

\title{Data science for urban equity: Making gentrification an accessible topic for data scientists, policymakers, and the community}

\numberofauthors{8}
\author{
\alignauthor
Bernease Herman\\
       \affaddr{University of Washington}\\
       \affaddr{Seattle, WA}\\
       \email{bernease@uw.edu}
\alignauthor
Gundula Proksch\\
       \affaddr{University of Washington}\\
       \affaddr{Seattle, WA}\\
       \email{prokschg@uw.edu}
\alignauthor
Rachel Berney\\
       \affaddr{University of Washington}\\
       \affaddr{Seattle, WA}\\
       \email{rberney@uw.edu}
\and
\alignauthor
Hillary Dawkins\\
       \affaddr{University of Washington}\\
       \affaddr{Seattle, WA}\\
       \email{hdawkins@uw.edu}
\alignauthor
Jacob Kovacs\\
       \affaddr{University of Washington}\\
       \affaddr{Seattle, WA}\\
       \email{kovjac19@uw.edu}
\alignauthor
Yahui Ma\\
       \affaddr{University of Washington}\\
       \affaddr{Seattle, WA}\\
       \email{maya16@uw.edu}
\and
\alignauthor
Jacob Rich\\
       \affaddr{University of Wisconsin}\\
       \affaddr{Madison, WI}\\
       \email{jrich3@wisc.edu}
\alignauthor
Amanda Tan\\
       \affaddr{University of Washington}\\
       \affaddr{Seattle, WA}\\
       \email{amandach@uw.edu}
}

\maketitle






\section{Introduction}
The University of Washington eScience Institute runs an annual Data Science for Social Good (DSSG) program that selects four projects each year to train students from a wide range of disciplines while helping community members execute social good projects, often with an urban focus \cite{Rokem15}. 

We present observations and deliberations of one such project, the DSSG 2017 `Equitable Futures' project, which investigates the ongoing gentrification process and the increasingly inequitable access to opportunities in Seattle. Similar processes can be observed in many major cities. The project connects issues usually analyzed in the disciplines of the built environment, geography, sociology, economics, social work and city governments with data science methodologies and visualizations. 

The project team developed a tool (Figure 1) that allows stakeholders in the city's development process to analyze, model, and visualize existing trends and the impact of potential changes in the built environment. The tool consists of two major parts, a (1) visualization tool and a (2) structural equation model. We've created interactive web mapping tool that visualizes equity indicators---primarily related to housing and development, income, mobility, and education---on the city and neighborhood scale. The structural equation model establishes and predicts relationships between publicly available data and underlying phenomena that cannot be measured directly. Furthermore, our chosen methodology allows us to both analyze biases introduced by aggregating correlated indicators and provide interpretable explanations of the underlying model. 

\begin{figure*}
\includegraphics[width=6.5in, keepaspectratio]{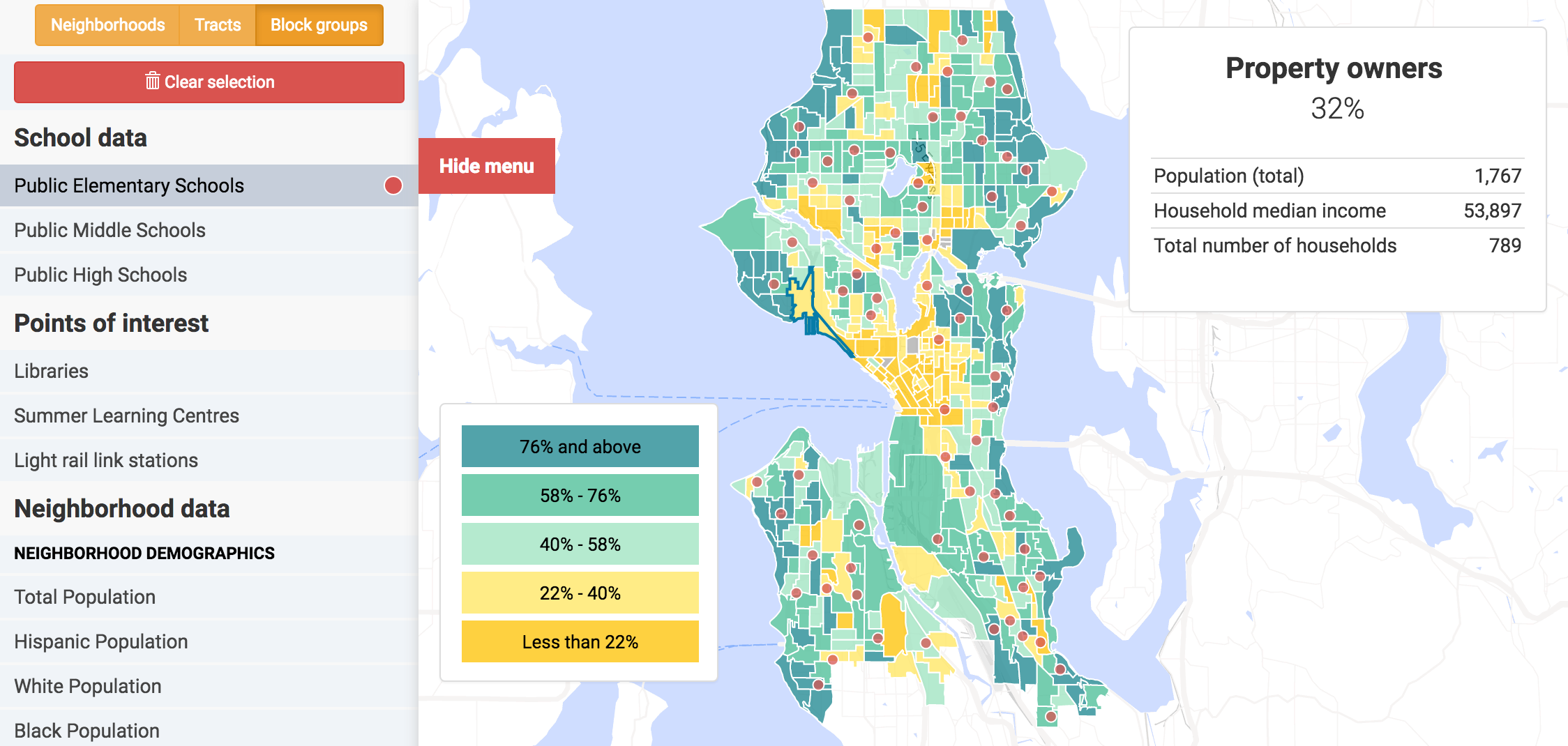}
\caption{One view of the \textit{Equitable Futures} visualization tool.}
\end{figure*}

\vfill\null

\section{Observations and Deliberations}
During the initial development phase of the project, we've collected insights on handling the interdisciplinary nature of the problem and maintaining interpretability while using advanced statistical techniques.

\subsection{Interdisciplinary problem specification}
Gentrification and inequity are inherently complex, multifaceted societal problems that various disciplines try to tackle with their own disciplinary methods. With no universal definitions of these processes at hand, various disciplines have tackled them with their own methods and a limited set of indicators and datasets. Built environment professionals, for example, employ mapping to understand the spatial distribution of indicators of gentrification and their change over time \cite{Chapple09}, aiming to create policies to manipulate the development of neighborhoods. Sociologists and economists describe these processes with statistics based on socioeconomic data \cite{Florida17}.

Our approach attempts to establish a more comprehensive definition of the problem, which combines the widely practiced method of equity or opportunity mapping with the statistical analysis of the data science community.

\subsection{Accessibility and collaboration across interdisciplinary teams}
In terms of collaboration, our interdisciplinary data science team that spans different disciplines and skill-levels must find ways to make complex domain knowledge accessible to all team members and external collaborators, in our case on a very short timeline. Our team has responded to the challenges of collaboration and accessibility by developing a literature review and educational workflow meant to support efficient division of labor while (1) building a conceptual common ground for the team (as a prerequisite for working critically with the data), and (2) laying the foundation for informational content to accompany, contextualize, and support critical use of the core interactive mapping tool. 

Unlike many urban planners and policy makers, who often have licenses to specialized proprietary software in their field, we follow a common data science practice of using open source software tools whenever possible. Not only has open source software been found to be less error-prone \cite{Synopsys15}, the use of open source projects in urban data projects helps to enable more civic participation and tool-making, while also lending transparency and access to the underlying implementation of the project dependencies.

\subsection{Algorithmic transparency and interpretability}
Social good demands that the methods and results of a data science project be both responsible and intelligible to various community audiences, enhancing their understanding of the domain and enabling them to have an accurate view of the assumptions and limitations of the project. Many established equity indices are weighted sums of indicators, without thorough analysis of the biases and redundancy of those model inputs. Our view is that methods like structural equation modeling allow us to communicate and mitigate the dangers of covariance between input variables, leading to less biased measures of equity.

We focus heavily on data transparency and model interpretability across the project. Data transparency extends beyond our use of publicly available data to include detailed records of our data preprocessing methodology. There is not yet an axiomatic definition of model interpretability \cite{Lipton17} but it is said that "interpretable" models engender trust in the producers and consumers of such models \cite{Kim15, Ribeiro16}. This is especially important in municipal decision making, where the output of the models inform high stakes decisions made by people who potentially lack the statistical background to fully understand the model assumptions and results \cite{Townsend14}.

\section{Acknowledgments}
The authors would like to thank the University of Washington eScience Institute and the University of Washington Runstad Center for Real Estate Studies. Many thanks to the DSSG student interns for their participation and hard work, and to the various mentors, speakers, and facilitators who graciously volunteered their time to contribute to the UW DSSG 2017 Program. This project was funded by generous grants to the eScience Institute from the Gordon and Betty Moore Foundation and the Alfred P. Sloan Foundation.

\nocite{*}
\bibliographystyle{abbrv}
\bibliography{references}

\begin{thebibliography}{1}

\bibitem{Chapple09}
K.~Chapple.
\newblock {\em Mapping Susceptibility to Gentrification: The Early Warning
  Toolkit}.
\newblock University of California, Berkley, CA, 2009.

\bibitem{Florida17}
R.~Florida.
\newblock {\em The New Urban Crisis: How Our Cities Are Increasing Inequality,
  Deepening Segregation, and Failing the Middle Class--and What We Can Do About
  It}.
\newblock Basic Books, New York, NY, 2017.

\bibitem{Kim15}
B.~Kim.
\newblock Interactive and interpretable machine learning models for human
  machine collaboration.
\newblock {\em PhD thesis, Massachusetts Institute of Technology}, 2015.

\bibitem{Lipton17}
Z.~C. Lipton.
\newblock ``the mythos of model interpretability.
\newblock {\em ICML}, 2017.

\bibitem{Ribeiro16}
M.~T. Ribeiro, S.~Singh, and C.~Guestrin.
\newblock ``why should i trust you?'' explaining the predictions of any
  classifier.
\newblock {\em KDD}, 2016.

\bibitem{Rokem15}
A.~Rokem et~al.
\newblock Building an urban data science summer program at the university of
  washington escience institute.
\newblock {\em The Bloomberg Data Science 4 Good Exchange}.

\bibitem{Synopsys15}
Synopsys.
\newblock Coverity® scan open source report 2014, 2015.

\bibitem{Townsend14}
A.~Townsend.
\newblock Legibility and interpretability in predictive models (of cities).

\end{thebibliography}

\end{document}